\documentclass{article}

\usepackage{arxiv}

\usepackage[utf8]{inputenc} 
\usepackage[T1]{fontenc}    
\usepackage{hyperref}       
\usepackage{url}            
\usepackage{booktabs}       
\usepackage{amsfonts}       
\usepackage{nicefrac}       
\usepackage{microtype}      
\usepackage{graphicx}
\usepackage{subfig}
\usepackage{float}
\usepackage{booktabs}
\usepackage{multirow}

\title{Statistical model of the human RF exposure in Small cells environment}

\author{
  A. Chobineh\\
  Telecom ParisTech, Chaire C2M \\
  Paris, France\\
  \texttt{amirreza.chobineh@telecom-paristech.fr} \\
   \And
  Y. Huang \\
  Telecom ParisTech, Chaire C2M \\
  Paris, France\\
   \And
   T. Mazloum \\
   Telecom ParisTech, Chaire C2M \\
   Paris, France\\
   \And
   E. Conil \\
   Agence nationale des fréquences (ANFR)\\
   Maisons-Alfort, France\\
   \And
   J. Wiart \\
   Telecom ParisTech, Chaire C2M \\
   Paris, France\\
}

\begin{document}
\maketitle

\begin{abstract}
Small cells are one of the solutions to face the imperative demand on increasing mobile data traffic. They are low-powered base stations installed close to the users to offer better network services and to deal with increased data traffic. In this paper, the global exposure induced in such networks as a whole from user equipment and base stations has been investigated. As the small cell is close to the user, the propagation channel becomes highly variable and strongly susceptible by environmental factors such as the road traffic. An innovative statistical path loss model is constructed based on measurements on two French commercial LTE small cells, operating at LTE 1800 MHz and 2600 MHz . This statistical path loss model is then used to assess global exposure of the adult proportion of a population in a scenario composed of a street lined with buildings, indoor and outdoor data users. 
\end{abstract}


\section{Introduction}
\label{intro}Nowadays radio frequency (RF) wireless communication systems are intensively used in different aspects of daily life. Phenomenal progresses in technologies, new applications, wireless devices and networks have led to a strong growth in data traffic over the last decade
\cite{Cisco} . To respond to such demand, research efforts have been carried out to improve the efficiency of communication protocols.
Current mobile networks are essentially made up of long-range radio equipment (macro cells) deployed to ensure coverage in the different territories. To face 5G ambitious aims (even more data traffic, higher frequency bands with poorer propagation capacities), this long-range layer will be supplemented by medium-range antennas (small cells) to improve coverage and connectivity for all, as the population increasingly uses mobile networks.

Small cells (SC) are low-powered base stations (BS). 
Deploying SCs results in offloading traffic from macro cells, hence, network performance, coverage, and capacity are increased. The coverage of a SC is less than a macro cell: it is ranged typically from a few meters to a few hundreds of meters\cite{SC}. 
Furthermore, SC antenna is often deployed at a low height, e.g., on top of a bus station. As a consequence, SC can be more numerous than macro cells and antennas closer to the users than macro antennas typically installed on a roof top or on top of a mast. 
By bringing the antennas closer to the users, it is also expecting to reduce the emitted power by user equipements (UE) and to increase the available throughput for users. 
UEs are therefore able to benefit from SCs, their greater quality and larger throughput than those of classical macro cells. SCs are expected to be deployed massively in order to cover the maximum number of users but such deployment can result to an important raise in public concerns. 
Despite the intensive use of wireless communication devices and existing international safety limits, there is still public concerns and risk perceptions about human exposure to RF electromagnetic fields (EMFs).
 The public debates on RF EMF human exposure often focus on the exposure induced by the cellular base stations and at large by the access points such as WiFi routers and Femto systems. 
As shown in works carried out within the European project called Low Electromagnetic Field Exposure Networks (LEXNET)\cite{Varsier} , a relevant complementary approach to assess the exposure is to deal with the global exposure in which both EM radiations coming from the UE (the ``uplink exposure'' ) and the infrastructure (the ``downlink exposure'' ) are taken into account.
As a matter of fact, the study carried in \cite{gati2010duality}  has proofed that there is an important correlation between EM radiations transmitted by UE and received from BS. However, for a long time the downlink and uplink exposures have been considered and assessed separately.
In the framework of LEXNET, a new exposure metric named Exposure Index (EI) was developed to quantify the exposure induced by UE and BS simultaneously \cite{Varsier}. 
Previous studies \cite{Varsier} \cite{Huang16} have computed EI for different scenarios involving urban LTE and UMTS networks. In these studies, a deterministic simulation tool was used to assess the path loss between BS  and the UE.
Such deterministic tool is strongly dependent on the level of details of buildings and terrain data. In addition, in these studies, the variability linked to propagation environment is not taken into account.\\
Recently, a novel statistical method was proposed \cite{Huang17} to cover the architectural variability by exploring the path loss exponents (PLE) observed in a typical city. Briefly, a distribution of PLE was given for a whole city via a stochastic geometry tool.
Different statistical path loss models were thus built based on the logarithm of the distance weighted by PLE and the path loss variability has been assessed through statistical methods for macro cell deployments.\\
 As the small cell is close to the user, the propagation channel is strongly dependent on the local environment and can be affected by any minor change, e.g., a passing car. It is therefore significant to investigate the variability of PLE for different UE located at different places instead of giving just one PLE distribution for an entire city as shown in the study \cite{Huang17}. In addition, as presented in the study\cite{Zeghnoun}, French people spend about 70\% of their time in indoor areas. Accordingly, indoor users stand for an important part of the population exposure which should be investigated. 
 
To overcome limits stated above, this paper presents the assessment of the global exposure induced in a LTE SC network based on a series of measurements performed using hand-held drive test mobile systems. The variability of the environment has been modeled through statistical PLE models for each individual LTE SC user located at different places, i.e., in indoor and outdoor areas. These innovative statistical models have been verified through the measurements.\\
The paper is  divided in  following sections. First, the EI metric, the adopted approach in statistical path loss modeling method as well as the assessment of uplink and downlink powers are described. Then, a scenario has been designed and the SC users has been assessed through the presented statistical methods. Finally, results of EI metric for indoor and outdoor users are presented. 
\section{Method and material }
In this section, the concept of Exposure Index (EI) has been presented. Then the measurement campaign, smallcells and measurement tools characteristics are introduced. Using the measurement results, a pathloss model has been constructed through the linear regression method. It is shown that due to the extreme variability of smallcell environment, this model suffers from high uncertainties since the standard deviation values are high. 
\\In order to model the high variation of smallcell environment, a statistical pathloss model has been proposed, based on measurement results. Through this model, we have computed the downlink received power from smallcell and the uplink emitted power by UE for each measurement point. Then the computed and measured results have been compared with each other. At the end of the chapter, it is shown that the proposed model has better accuracy than the one constructed through the linear regression method.
\label{sec:1}
\subsection{Population exposure and Exposure Index }
\label{sec:2}
The Exposure Index (EI) quantifies the whole or partial body global exposure (induced by UE and infrastructures), averaged over time and over a population in a given geographical area.  EI metric aggregates the exposure  by both the near-field and the far-field sources. For a given geographical area, EI can be described as the following \cite{Varsier}:
\begin{equation} \label{eq:1}
EI^{SAR} = \frac{1}{T}\sum_{t}^{N_{T}}\sum_{p}^{N_{P}}\sum_{e}^{N_{E}}\sum_{r}^{N_{R}}\sum_{c}^{N_{C}}\sum_{l}^{N_{L}}\sum_{pos}^{N_{pos}}f_{t,p,e,r,l,c,pos}\left ( \sum_{u}^{N_{U}}(d^{UL}\;\overline{P}_{T_X})+d^{DL}\;\overline{S}_{RXinc}\right )
\end{equation}
Where : 
\begin{itemize}
  \item $EI^{SAR}$ is the Exposure Index value, the average exposure of the population of the considered geographical area  over the considered time frame T. SAR refers to whole-body SAR, organ-specific SAR or localized SAR.
  \item $N_T$ is the number of considered Time periods of the day
  \item $N_P$ is the number of considered Population categories
  \item $N_E$ is the number of considered Environments
  \item $N_R$ is the number of considered Radio Access Technologies (RAT)
  \item $N_C$ is the number of considered Cell types
  \item $N_L$ is the number of considered Load profiles
  \item $N_{Pos}$ is the number of considered Postures 
  \item $N_U$ is the number of considered Usages with devices
\end{itemize}
EI takes into consideration different exposure sources and exposure situations. The formulation of EI contains a set of different parameters such as different time periods (t), different population categories (p), different environments (e),  network technologies (r),cell types (c), different load profiles (l), body postures (pos) and usages (u). 
$\overline{P}_{TX}$ is the mean $TX$ power transmitted by UE during uplink usage period. $\overline{S}_{RXinc}$ represents the mean incident power density on the human body during period t. \\The coefficient $d^{UL}$ is the uplink exposure induced by UE transmissions. It is expressed as an absorbed dose normalized to a transmitted power of 1 W : 
\begin{equation} \label{eq:dul}
d^{UL}_{\frac{s}{kg}} = \frac{TD^{UL}_{t,p,l,e,r,c,u,pos[s]}SAR^{UL}_{p,r,u,pos[\frac{W}{kg}]}}{P^{ref}_{TX[W]}}
\end{equation}
Where:
\begin{itemize}
  \item $TD^{UL}_{t,p,l,e,r,c,u,pos[s]}$ is the time duration of usage u, and a user profile l, when connected to the RAT r, operating in cell type c, in environment e, for the population category p, in the posture pos, during the time period t. 
  \item $\frac{SAR^{UL}_{p,r,u,pos[\frac{W}{kg}]}}{P^{ref}_{TX[W]}}$ can be the whole body or an organ-specific or tissue-specific SAR value for the usage u and the posture pos, in the frequency band of RAT r, and the population category p, calculated for an incident emitted power of $P^{ref}_{TX[W]}$ and normalized to this power. 
\end{itemize}  
The coefficient $d^{DL}$ is associated to the exposure induced by the downlink. It is expressed as an absorbed dose normalized to an incident power density of $1W/m^2$.
\begin{equation} \label{eq:ddl}
d^{DL}_{\frac{s}{kg}} = \frac{TD^{DL}_{t,p,e,r,c,pos[s]}SAR^{DL}_{p,r,pos[\frac{W}{kg}]}}{S^{ref}_{RX[W]inc}}
\end{equation}
Where:
\begin{itemize}
  \item $TD^{DL}_{t,p,e,r,c,pos[s]}$ is the time duration of posture pos, when connected to the RAT r, operating in cell type c, in the environment e, for the population p, during the time period of the day t.  
  \item $ \frac{SAR^{DL}_{p,r,pos[\frac{W}{kg}]}}{S^{ref}_{RX[W]inc}}$ can be the whole body or an organ-specific or tissue-specific SAR value induced by the base station or access points of RAT r, in the population p, for the posture pos, normalized to the power density $S^{ref}_{RX[W]inc}$.
\end{itemize} 
Finally the coefficient $f_{t,p,e,r,l,c,pos}$ is the fraction of the population p, with user profile l, in posture pos connected to RAT r for cell type c in environment e during time t. 
In this paper, EI is evaluated for adults using data service in standing position connected to a LTE SC at 1800 or 2600 MHz frequencies in an urban environment.  
In order to assess the EI, the main difficulty is to estimate the actual level of emitted and received power by UE and upload data rate. 
It is worth mentioning that the user is exposed to downlink emitted power from the SC permanently and constantly. The uplink exposure time, in other words, the uplink traffic transmission time depends on  the throughput and the volume of uploaded data traffic. 

In our study, emitted and received power, throughput and uplink power control algorithm parameters were assessed through  a series of drive-test measurements on two SC sites. By measuring these parameters while traveling in the line of sight (LoS) direction of a SC, the variability of received and emitted powers by the UE can be thus characterized. 
The configuration of measured SCs is presented in the next part.

\subsection{Small cells configuration}
\label{sec:2}

\begin{figure}[h]
 \centering
  \includegraphics[width=0.45\textwidth]{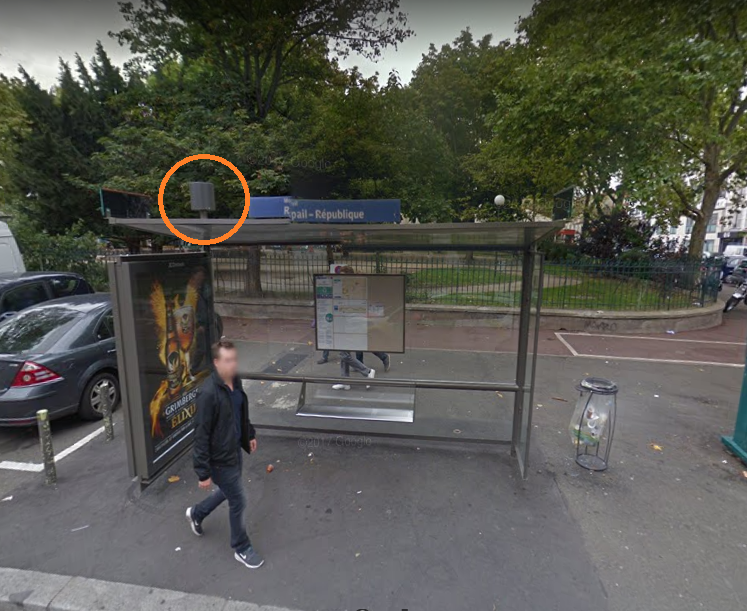}
\caption{SC deployed on top of a bus stop}
\label{fig:2}       
\end{figure}

SC antennas are usually deployed on street furniture (e.g bus stops, billboards and streetlights) or building facades to increase the quality and the connectivity of cellular networks \cite{SC}. 
Recently, in France, trials have been conducted regarding to the SC deployment exposure \cite{Mazloum2}\cite{Mazloum1}. During these trials we had the opportunity to perform a series of measurements on two different SCs installed on top of advertisement panels and bus stops in a dense urban city (Figure \ref{fig:2}).
The streets width in which SCs are located  is about 8 m and average buildings height surrounding the sites is about 10 m. SC's characteristics are presented in Table 1.\\
The SCs used in this experiment are equipped with directive antennas having 38 dBm as equivalent isotropic radiated power (EIRP). Two LTE frequency bands are activated: 1800 MHz with 20 MHz of bandwidth and 2600 MHz with 15 MHz of bandwidth.

\begin{table}[]
\centering
\caption{Small cell characteristics}
\label{table1}
\begin{tabular}{|c|c|c|}
\hline
Cell identity  & 1             & 2                     \\ \hline
Bands & \multicolumn{2}{c|}{4G 1800, 4G 2600} \\ \hline
Height         & \multicolumn{2}{c|}{2.9 m}               \\ \hline
Mast type      & Bus station   & Advertisement panel   \\ \hline
\end{tabular}
\end{table}
\begin{figure}[h]
 \centering
  \includegraphics[width=0.45\textwidth]{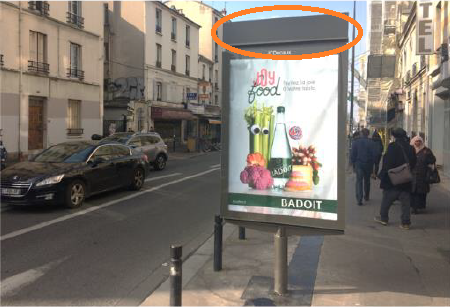}
\caption{SC deployed on top of an advertisement panel}
\label{fig:1}       
\end{figure}

\subsection{Measurement method and uplink and downlink power assessment}
\label{sec:2}
To assess the uplink and downlink exposure, it is essential to evaluate emitted and received power by the UE.
The downlink received power by UE can be presented by the following equation:
\begin{equation} \label{eq:2}
P_{R_x}=P_{T_x}+Gain_{SC}(\theta,\phi)- PL +Gain_{UE}(\theta',\phi')
\end{equation}
$PL$ denotes the channel path loss, $P_{T_x}$ the transmitted power by the SC, $Gain_{UE} $ the UE antenna gain, $Gain_{SC}$ the SC antenna gain when UE is place at the direction $(\theta,\phi)$. $P_{R_x}$ is the received power by UE. In this study the UE antenna is assumed to be omnidirectional ($Gain_{UE}(\theta',\phi')_{UE} = 0$).

The reference emitted power by SC and the received power by UE are named in LTE protocol respectively as Reference Signal (RS) and  Reference Signal Received Power(RSRP).   
Thus, the received power by the UE can be assessed through the following equation: 
\begin{equation} \label{eq:3}
RSRP = RS + Gain_{SC}(\theta,\phi) - PL
\end{equation}
In LTE technology the emitted power by UE is determined through an uplink power control algorithm.
The power control algorithm is described as follows~\cite{3GPP}~: 
\begin{equation} \label{eq:puplink}
P_{T_x}=min(P_{max},10\;log_{10}(M)+P_0+\alpha\; PL+\delta_{TF}+f_g)
\end{equation}
This algorithm is a combination of open and closed loop components. The open loop power control (OLPC) is responsible for a rough setting of uplink transmitted power \cite{gora2010}.
It compensates slow changes of path loss in order to achieve a target power $P_{0}$ for one resource block.
Then the total number of allocated uplink resources block (denoted M) is added in decibels to this power.
The path loss ($PL$) is compensated by a factor $\alpha$ which can be fixed at one of the [0,0.4,0.5,0.6,0.7,0.8,0.9,1] values.
When $\alpha$ is 1, the measured path loss is fully compensated in uplink transmissions. 
For $\alpha$ = 0, path loss between UE and BTS is ignored in power control algorithm and the emitted power is constant for all users. 
For $\alpha$ between 0.4 and 1, path loss compensation would be fractional. 
The closed loop power control (CLPC) component considers fast variations of propagation channel by use of a corrector ($f_g$ ) and also uplink modulation scheme ($\delta_{TF}$). Since the closed loop power control has a minor effect on emitted power compared to OLPC, it has been ignored in this study. $P_{max}$ is the maximum uplink transmit power in LTE and is equal to 23dBm ($\pm$ 2dBm). 
The need to record RS, RSRP, $P_{TX}$, and OLPC parameters to compute path loss motivates us to use drive test mobile hand-set technology as the most relevant solution.
A series of measurements using drive test mobile equipment has been carried out during trial in order to record the network characteristics such as throughput, emitted and received power by UE and OLPC parameters. These measurements have been performed in LoS direction of the SCs. The journey consisted of a round trip of about 200 meters.

To increase the set of data, two different drive test solutions have been used. By doing so, different antenna characteristics and efficiency of mobiles are considered \cite{Mazloum2}.
 Regarding the uplink emitted power from UE in active mode, the drive test hand sets have been programmed to upload 100 MBytes files on a FTP server repetitively. 
This configuration imitates a real user using data services on his mobile phone.\\
Hand-held drive test solutions used in this work are on the one hand Viavi JDSU\cite{jdsu} solution installed on Samsung Galaxy S4 and on the other hand  AZENQOS (AZQ)\cite{azq}, developed by Freewill FX Company Limited and installed on LG Nexus 5X.
The received signal strength, emitted signal by UE, GPS coordinates of each measurement point and uplink power control algorithm parameters have been measured and recorded. It is worth mentioning that at post processing stage, all the measurements have been merged and processed together. These parameters have been used in  model construction and validation process described in \ref{sec:stat_pathloss_24}.

\subsection{Statistical path loss modeling  }
\label{sec:stat_pathloss_24}
\subsubsection{Path loss modeling  } \label{PLmodeling}
\label{sec:3}
In this study, path loss model was constructed, based on measurements, according to the following equation \cite{Huang16}\cite{mobile_book}:
\begin{equation} \label{eq:PL}
PL=A+10\;\gamma\;log_{10}(\frac{d}{d_0})
\end{equation} 

Where $\gamma$ is the path loss exponent, A is the decibel path loss at the reference distance $d_0$ of 1 m and is computed as following : 

\begin{equation} \label{eq:5}
A=20\;log_{10}(\frac{4\pi}{\lambda})
\end{equation}

\begin{figure}[H] 
 \centering
  \includegraphics[width=0.75\textwidth]{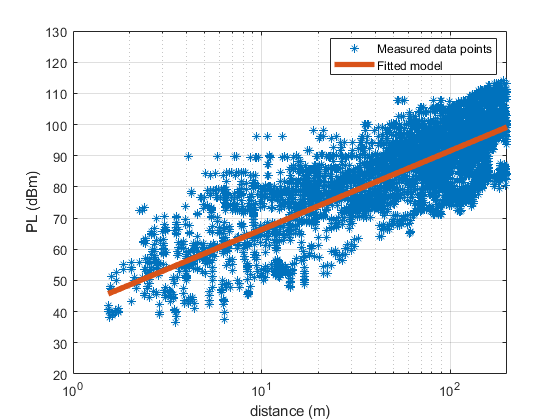}
\caption{Fitted model to 2600 MHz measured path loss data}
\label{fig:3}       
\end{figure}

In which $\lambda$ is  the wavelength of the signal.
On a log-log plot, for the scattered data, a linear curve has been fitted using linear regression method.
Figure \ref{fig:3} is a plot of the measured path loss and the fitted model in decibels in terms of distance between SC and UE at 2600 MHz.
Table 2 summarizes the $PL$ models characteristics obtained by the linear regression at both frequencies. 

\begin{table}
\centering
\caption{Path loss model's parameters}
\label{tab:2}       
\begin{tabular}{lllll}
\hline\noalign{\smallskip}
Frequency (MHz) & $\gamma$ & A(dB) & Mean difference(dB) & Standard deviation(dB) \\
\noalign{\smallskip}\hline\noalign{\smallskip}
1800 & 2.85 & 37 & 0.5 & 5.7 \\
2600 & 2.52 & 41 & 0.1 & 7.1 \\
\noalign{\smallskip}\hline
\end{tabular}
\end{table}
Mean differences between measured data and proposed models reveal that the proposed models are a good representation of data. However the standard deviation values are quite high which led us to construct a statistical path loss model.
\subsubsection{Statistical path loss modeling  }
\label{sec:statPL}
\begin{figure}[h] 
 \centering
  \includegraphics[width=0.75\textwidth]{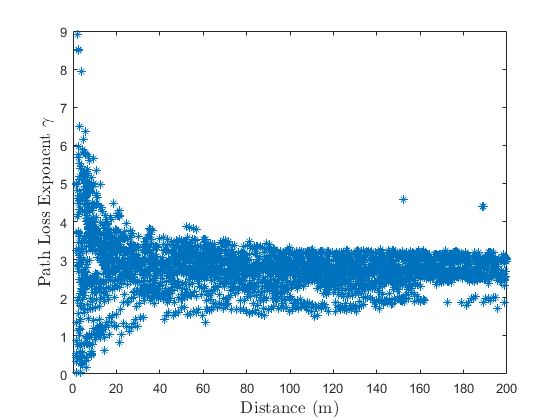}
\caption{The variation of path loss exponent ($\gamma$) in terms of distance at 2600~MHz.}
\label{fig:4}       
\end{figure}

\begin{figure}[h]
 \centering
  \includegraphics[width=0.75\textwidth]{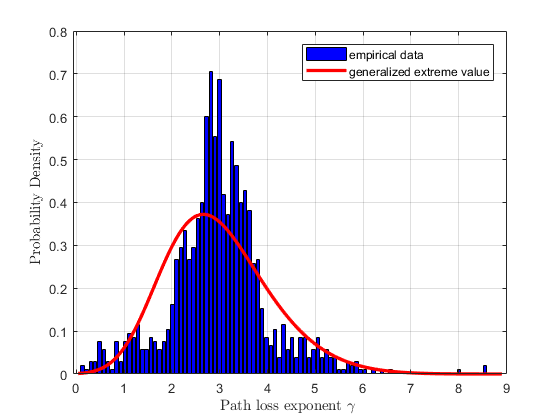}
\caption{Path loss exponent ($\gamma$) distribution for distances less than 60 m at 2600 MHz}
\label{fig:5}       
\end{figure}

\begin{figure}[H]
 \centering
  \includegraphics[width=0.75\textwidth]{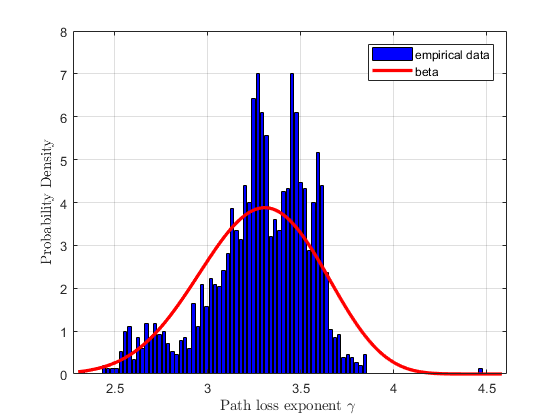}
\caption{Path loss exponent ($\gamma$) for distances larger than 60 m at 2600 MHz}
\label{fig:6}       
\end{figure}

The main challenge in path loss modeling is to consider the attenuation and variations in the propagation environment.
Since SCs are installed at little heights, environmental variations such as moving cars and pedestrians could seriously affect the propagation channel conditions even more than for usual macro cells networks.
For instance, a bus could completely block the SCs antenna. 
Therefore, the environment becomes highly variable specially for the distances close to the antenna. This variability of the environment implies that at the same distance from SC, in different times, different path loss values have been measured as it is shown in figure \ref{fig:3}.
In such environment, using usual deterministic approaches (such as the linear regression method applied in section \ref{PLmodeling}) to model the path loss, introduces important uncertainties in the final results. 
To take into account these variablities, the path loss exponent ($\gamma$) has been modeled through a statistical approach as presented in \cite{Huang17}.
This leads to predict the path loss at a specific distance not as one value but as a statistical distribution.
Consequently this model presents all the possible variations in the SC-UE channel in an urban area in LoS direction of a SC antenna. 
The following equation is used to assess the path loss exponent at each measurement point \cite{Huang17}:
\begin{equation} \label{eq:6}
\gamma =\frac{PL-A}{10\;log_{10}(\frac{d}{d_0})} 
\end{equation}
The value of $\gamma$ has been computed based on about 20000 measurement points on two sites at 1800 and 2600 MHz. Results on figure \ref{fig:4} show that $\gamma$ varies widely at close distances and it becomes more stable, for further distances. Hence, it is possible to classify two different $\gamma$ models depending on distance. According to measurements and using the Kolmogorov Smirnov test, for distances less than 60~m from the SC, $\gamma$ follows a generalized extreme value (GEV) distribution and for distances larger than 60~m, it follows a Beta distribution. These distributions were further used as inputs to assess the emitted and received power by user. Figure \ref{fig:5} and \ref{fig:6} illustrate the distributions of $\gamma$ at 2600~MHz for distances lower and higher than 60 m from the SC. The characteristics of these distributions are presented in table \ref{tab:gamma_char}.

\begin{table}[h]
\centering
\caption{Characteristics of the $\gamma$ distribution for the distances lower and higher than 60 m}
\label{tab:gamma_char}
\begin{tabular}{@{}ccccccc@{}}
\toprule
Frequency & Distance 	      & Distribution & Parmeter 1   & Parmeter 2   & Parmeter 3 & Parmeter 4 \\ \midrule
1800      & r\textless60 m    & GEV          & k=-0.31      & s=0.42       & m=2.7      &            \\
1800      & r\textgreater60m  & Beta         & $\alpha$1=3      & $\alpha$2=3.4       & a=2.2     & b=3.2      \\
2600      & r\textless60m     & GEV          & k=-0.23      & s=0.93       & m=2.6      &            \\
2600      & r\textgreater60m  & Beta         & $\alpha$1=21 & $\alpha$2=18 & a = 0           &  b = 5     \\ \bottomrule
\end{tabular}
\end{table}
These distributions are going to be used to compute the received power by UE by using equation \ref{eq:3} and \ref{eq:PL} for each measurement point and uplink emitted power by using equation \ref{eq:puplink}.
\subsection{Uplink emitted and Downlink received powers}
\label{sec:ul_dl}
\subsubsection{Downlink received power}
\label{sec:5}
According to the equation \ref{eq:3}, the received power by user can be assessed through the following formula :
\begin{equation} \label{eq:RSRP}
RSRP = RS + Gain_{SC}(\theta,\phi) -PL
\end{equation}

\begin{figure}[H]
 \centering
  \includegraphics[width=0.75\textwidth]{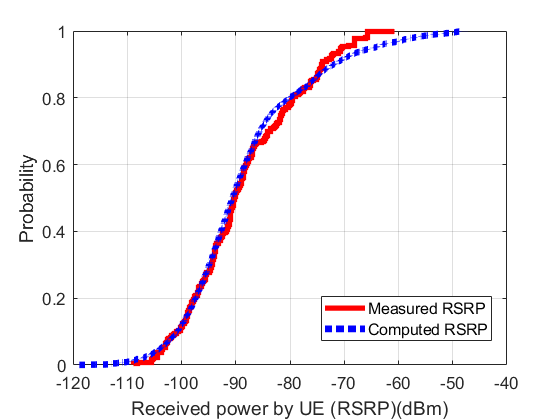}
\caption{Measured and computed received power CDF (RSRP) (dBm) at 2600MHz }
\label{fig:9}       
\end{figure}

For each measurement point, the statistical distribution corresponding to the distance between the point and the SC defined in \ref{sec:statPL} is used to sample the $\gamma$. Then, path loss value is computed for these point by equation \ref{eq:PL}.  Finally, the received power by UE at this point has been computed using equation \ref{eq:RSRP}. This process has been done for all the measurement points. 
The distribution of computed and measured RSRP have been presented in figure \ref{fig:9}. Table \ref{tab:rsrp} gives the main characteristics of measured and computed RSRP.  

\begin{table}[]
\centering
\caption{Characteristics of measured and computed RSRP}
\label{tab:rsrp}
\begin{tabular}{cccc}
\hline
RSRP         & Mean(dBm)   & Median(dBm) & Quantile 95\%(dBm) \\ \hline
Measured & -88.6  & -90.62 & -65.91        \\
Computed & -88.55 & -90.14 & -70.26       
\end{tabular}
\end{table}
\clearpage
\subsubsection{Uplink emitted power}
\label{sec:5}
\begin{figure}[H]
 \centering
  \includegraphics[width=0.75\textwidth]{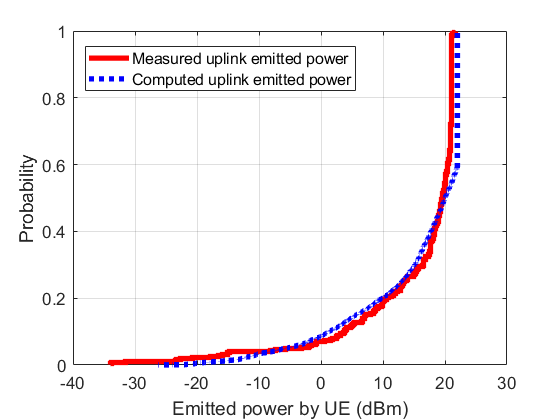}
\caption{Mesured and computed uplink emitted power CDF at 2600 MHz (dBm)}
\label{fig:emittedP}       
\end{figure}

According to equation \ref{eq:puplink} and by ignoring the closed loop component of the uplink power control algorithm, the uplink emitted power can be assessed through the following formula: 
\begin{equation} \label{eq:8}
P_{T_x} = min(P_{max},10 \;log_{10}(M)+P_0+\alpha\;PL)
\end{equation}
 The drive test mobile equipment records the number of uplink assigned resources block(M) for each time slot.
The measured $\alpha$ shows that this parameter is constant and fixed to 1 at all time for SCs: the path loss is fully compensated by the uplink power control algorithm.
Based on recorded values, $P_0$ value has been fixed at -96 dBm.  
The path loss ($PL$) parameter has been computed by using statistical models determined in \ref{sec:statPL}. Figure \ref{fig:emittedP}, compares the distribution of computed uplink emitted powers and measured uplink emitted powers. Table \ref{tab:char_uplink} gives the characteristics of measured and predicted emitted powers. 
Results show that the distribution of emitted and received powers by UE can be estimated through these models with a good precision. 

\begin{table}[H]
\centering
\caption{Characteristics of measured and computed uplink emitted power}
\label{tab:char_uplink}
\begin{tabular}{cccc}
\hline
Uplink emitted power         & Mean (dBm)  & Median(dBm) & Quantile 95\%(dBm) \\ \hline
Measured & 15.24  & 19.54 & 21.01        \\
Computed & 15.8 & 20.27 & 22      
\end{tabular}
\end{table}

\clearpage

\section{LTE SC scenario simulation}
\subsection{Scenario description}
\begin{figure}[h]
 \centering
  \includegraphics[width=0.75\textwidth]{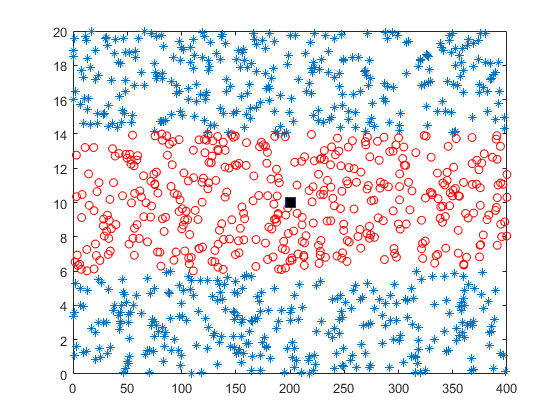}
\caption{Scenario structure : indoor observations ( $\ast$ ) and outdoor observations ($\circ$). The SC is placed at the center of the map. Street length : 400m and street width : 8m }
\label{fig:10}       
\end{figure}

\begin{figure}[h!]
 \centering
  \includegraphics[width=0.75\textwidth]{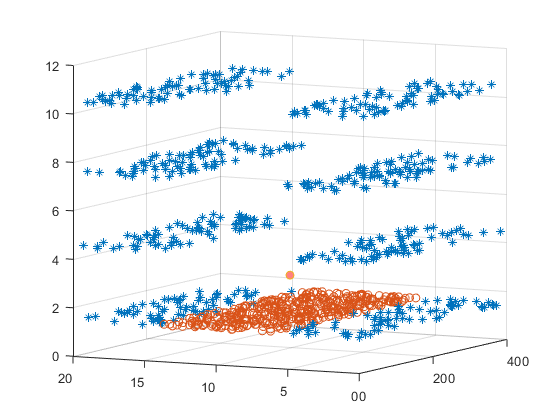}
\caption{Scenario structure(3D)  for indoor(blue $\ast$) and outdoor(red $\circ$) observations}
\label{fig:11}       
\end{figure}

In order to assess the exposure of a population using an LTE SC in indoor and outdoor areas, 
a simplistic scenario has been designed as a street of 8 m width and 400 m length. Two series of buildings are considered at each side of the street. The penetration depth is considered to be 6 m in buildings. Each building has 4 floors and the height of each floor is 3 m. The height of the UE from the ground is 1.5 m. Due to the lack of information on the architecture of indoor environment, the penetration loss has been taken into account as a random variable uniformly distributed on the interval [7-13] dB. Figure \ref{fig:10} and \ref{fig:11} illustrate the structure of the scenario. 
A sample $10^5$ different random observations has been constructed in order to cover user's positions. Users are distributed uniformly in the indoor and outdoor areas. Since people spend about 70\% of their time in indoor we have considered that 70\% of the occurrences happen in the indoor area \cite{Zeghnoun}.
The SC is placed in the center of the map in the outdoor area at 3 m from the ground.
For each observation, we have computed the distance between user and the SC. Then random sampling of the appropriate distribution of path loss distribution (GEV or Beta according to distance) has been carried out in order to assess a path loss distribution for each observation.
Having the path loss distribution, we can assess the distribution of received and emitted powers for each observation as described in \ref{sec:ul_dl}. 
Simulation configurations are presented in table \ref{network}. 
The EI was evaluated over 1h time frame for LTE SC users in dense urban area. Only adult users and moderate data usage have been taken into account. The scenario includes indoor and outdoor users. Considering future increased data usage,
it is assumed that in average each user upload 100 MBytes of data through the SC each day. This means that during 1h time frame, each user uploads on average 4.16  MBytes. Based on the received power by user, the Signal to Noise ratio (SNR) and finally the throughput of each user has been assessed. The uplink emission time (time taking by a user to upload 4.16 MBytes of data to the network) can be assessed by dividing the data volume by the throughput.   
\begin{table}[h]
\centering
\caption{Network parameters for LTE SC in urban area}
\label{network}
\begin{tabular}{lll}
\hline
Environment                     & \multicolumn{2}{l}{Typical European dense urban}                                                                                      \\ \hline
\multirow{8}{*}{System}         & Cell type                                           & Small cell                                                                      \\
                                & Carrier                                             & LTE                                                                           \\
                                & Carrier frequency                                   & 1800 MHz, 2600 MHz                                                              \\
                                & Antenna                                             & Omni-directional with 0 dBi gain                                                \\
                                & Reference Signal                                    & 10 dBm                                                                          \\
                                & Bandwidth                                           & \begin{tabular}[c]{@{}l@{}}20 MHz at 1800 MHz\\ 15 MHz at 2600 MHz\end{tabular} \\
                                & Thermal noise                                        & -101 dBm                                                                        \\
                                & Antenna height                                      & 3 m                                                                             \\ \hline
\multirow{5}{*}{User equipment} & Max/Min transmission                                & 23 dBm/-40dBm                                                                   \\
                                & Antenna                                             & Omni-directional with 0 dBi gain                                                \\
                                & Penetration loss  Indoor${\leftrightarrow}$ Outdoor & 10$\pm$3 dB                                                                     \\
                                & Upload file volume                                  & 4.16 MB per hour                                                                \\
                                & UE height                                           & 1.5 m from the ground                                                           \\ \hline
\end{tabular}
\end{table}

\begin{table}[]
\centering
\caption{Exposure Configuration}
\label{EI}

\begin{tabular}{cc}
\hline
Time period (t)  & 1 h                                                                    \\
RAT (r)          & LTE                                                                    \\
Cell type (c)    & Small cell                                                             \\
Environment (e)  & \begin{tabular}[c]{@{}c@{}}Urbain city\\ Indoor\\ Outdoor\end{tabular} \\
Population (p)   & Adult                                                                  \\
User profile (l) & \begin{tabular}[c]{@{}c@{}}Moderate\\ (100 MB upload per day)\end{tabular}          \\                                      
Usage (u)        & Data                                                                   \\
Posture (pos)    & Standing                                                               \\ \hline
\end{tabular}
\end{table}


\subsection{Exposure}
\begin{figure}[H]
 \centering
  \includegraphics[width=0.75\textwidth]{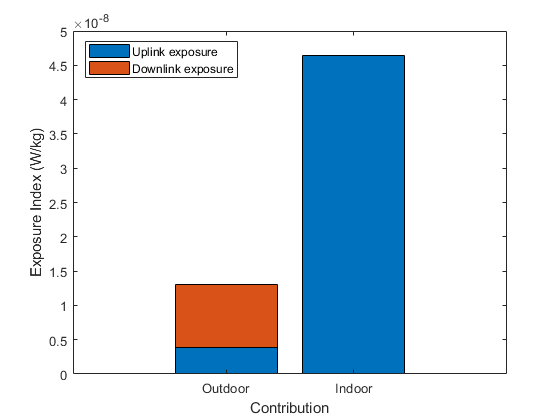}
\caption{Exposure Index of indoor and outdoor users and contribution of uplink and downlink exposure}
\label{fig:12}       
\end{figure}

Once the average values of emitted and received powers of all the observations are computed, it is possible to assess the EI by using normalized whole-body SAR values. 
The whole body reference SAR values induced by mobile phones and base stations of adult for data usage were extracted from the values computed in the framework of LEXNET project.   
EI configurations are presented in table \ref{EI} and table \ref{tab:EI_freq} summarizes EI results obtained at 1800 and 2600 MHz.
\begin{table}[h]
\centering
\caption{EI at different frequencies}
\label{tab:EI_freq}       
\begin{tabular}{ccc}
\hline
\multirow{2}{*}{Parameters}     & \multicolumn{2}{c}{Frequencies}            \\ \cline{2-3} 
                                & 1800                & 2600                 \\ \hline
$\overline{P}_{T_X}$(W)         & $7.2\times10^{-3}$  & $7.5\times10^{-3}$   \\
$SAR^{UL}_{p,r,u,pos}$(W/kg)    & 0.0039              & 0.0029               \\
$\overline{S}_{RXinc}$($W/m^2$) & $8.91^{-10}$         & $9.36\times10^{-10}$ \\
$SAR^{DL}_{p,r,pos}(W/kg)$      & 0.0047              & 0.0042               \\
Downlink exposure (W/kg)        & $1.89\times10^{-9}$ & $3.71\times10^{-9}$  \\
Uplink exposure (W/kg)          & $3.43\times10^{-8}$ & $2.66\times10^{-8}$  \\
EI (W/kg)                       & $3.62\times10^{-8}$ & $3.03\times10^{-8}$  \\ \hline
\end{tabular}
\end{table}
Table \ref{tab:in_out} presents the results at 2600 MHz for indoor and outdoor users.
The results have shown that, in this configuration, the public global exposure of indoor LTE SC users is about 4 times higher than the outdoor users (figure \ref{fig:12}). This is due to higher emitted power by UEs in indoor compared to outdoor. 
These results, highlight the importance of the uplink exposure in global exposure and EI assessment. 
\begin{table}[h]
\centering
\caption{EI comparison between indoor and outdoor exposure of users at 2600 MHz}
\label{tab:in_out}  
\begin{tabular}{ccc}
\hline
\multirow{2}{*}{Parameters}     & \multicolumn{2}{c}{Environment}            \\ \cline{2-3} 
                                & Outdoor             & Indoor               \\ \hline
$\overline{P}_{T_X}$(W)         & $1.3\times10^{-3}$  & $1.16\times10^{-2}$  \\
$SAR^{UL}_{p,r,u,pos}$(W/kg)    & 0.0029              & 0.0029               \\
$\overline{S}_{RXinc}$($W/m^2$) & $2.32^{-9}$         & $7.6\times10^{-12}$  \\
$SAR^{DL}_{p,r,pos}(W/kg)$      & 0.0042              & 0.0042               \\
Downlink exposure (W/kg)        & $3.84\times10^{-9}$ & $3.01\times10^{-11}$ \\
Uplink exposure (W/kg)          & $9.23\times10^{-9}$ & $4.63\times10^{-8}$  \\
EI (W/kg)                       & $1.30\times10^{-8}$ & $4.64\times10^{-8}$  \\ \hline
\end{tabular}
\end{table}

\clearpage
\section{Conclusion}
Small cells appear as a solution for future mobile networks. 
Being closer to the user, SCs can provide higher network performance but they can also be influenced by local environment variations. 
Being more numerous, SCs attract attention in terms of induced EMF exposure, the assessment of such exposure is therefore important.
In this paper, the great variability of local propagation channel in case of SC is taken into account by using an innovative statistical approach. RF exposure is evaluated as a whole by using the new metric called EI, introduced by LEXNET project. Innovative statistical path loss models have been built using the measurements performed with SCs installed in a French urban city. EI has been computed by using these models in an indoor and outdoor scenario.
In this configuration, results show that the uplink component of exposure is  as expected dominating EI and the exposure of indoor users is higher than outdoor users due to the contribution of uplink exposure. In addition the computed EI value for the presented scenario is  $3.62\times10^{-8} W/kg$ at 1800 MHz and $3.03\times10^{-8} W/kg$ at 2600 MHz which is less than values reported in UMTS and LTE macro cell scenarios in other studies \cite{Huang17}\cite{Huang16}.
This can be explained by the reduced distance between SC and UEs and therefore the reduced uplink emitted powers. 
In this study the designed scenario is quite simplistic. More complex configurations will be investigated in future works. 
 
\section*{Acknowledgement}
This research has been performed in the framework of AMPERE. It was supported by ANSES (Agence Nationale S\'ecurit\'e Sanitaire Alimentaire Nationale(EST-16-RF-04)) and ANFR (Agence nationale des fr\'equences (convetion 83 ANFR 2016)) .

\end{document}